\documentclass[aps,eqsecnum,prd,twocolumn]{revtex4}
\usepackage{graphics,graphicx}
\usepackage{amsmath}
\usepackage{amssymb,latexsym,mathrsfs}
\usepackage{hyperref}

\def\bea{\begin{eqnarray}}
\def\eea{\end{eqnarray}}
\def\ba{\begin{array}}
\def\ea{\end{array}}

\def\beq{\begin{equation}}
\def\eeq{\end{equation}}

\begin{document}

\title{Weak value of Dwell time for Quantum Dissipative spin-1/2 System}

\author{Samyadeb Bhattacharya$^{1}$ \footnote{sbh.phys@gmail.com}, Sisir Roy$^{2} $ \footnote{sisir@isical.ac.in}}
\affiliation{$^{1,2}$Physics and Applied Mathematics Unit, Indian Statistical Institute, 203 B.T. Road, Kolkata 700 108, India \\}

\vspace{2cm}
\begin{abstract}

\vspace{1cm}

\noindent The dwell time is calculated within the framework of time-dependent weak measurement considering dissipative interaction between a spin-$\frac{1}{2}$ system and the environment. Caldirola and Montaldi's method of retarded Schr\"{o}dinger equation is used to study the dissipative system. The result shows that inclusion of dissipative interaction prevents the zero time tunneling.

\vspace{2cm}


\vspace{1cm}

\end{abstract}

\vspace{1cm}

\maketitle

\section{Introduction}
The recent experimental results \cite{1} on superluminal tunneling speed raise lot of controversy among the community.
The problem of defining tunneling times has a long history \cite{1a},simultaneously with the fundamental problem of introducing time as a quantum mechanical observable and, in particular, of a definition (in Quantum Mechanics) of the collision durations.
In fact,experiments  on transmitting information containing features of an optical pulse across the ``fast light" medium, in which the group velocity exceeds the vacuum speed of light c, have renewed the interest in the so called ``superluminal" propagation phenomena. This superluminality have been predicted in connection with various quantum systems propagating in a forbidden zone. Aharonov et.al \cite{2} along with other authors \cite{2a} dealt with the problem of tunneling time from the context of weak measurement.The notion of the weak value of a quantum mechanical observable was originally  introduced by Aharanov et.al \cite{3}-\cite{5}. This quantity is the statistical result of a standard measurement procedure performed upon a pre selected and post selected (PPS) ensemble of quantum systems when the interaction between the measurement apparatus and each system is sufficiently weak. Unlike the standard strong measurement of a quantum mechanical observable which sufficiently disturbs the measurement system, a weak measurement of an observable for a PPS system does not appreciably disturb the quantum system and yields the weak value as the measured value of the observable. Aharanov et.al \cite{2} have shown that in their approach , tunneling time corresponds to superluminal velocity. On the other hand, experiments with photonic band gap structures \cite{5a,5b} also showed apparent superluminal barrier traversal. These observations as well as the theoritical predictions lead towards the phenomena of superluminal barrier traversal. in case of superluminality, Winful suggested \cite{5c,5d,5e,5e1,5e2} an explanation of faster than light phenomena by the concept of energy storage and release in the barrier region. He argued that the group delay, which is directly related to the dwell time with an additive self interaction delay \cite{5f}, is actually the lifetime of stored energy (or stored particles) leaking through both ends of the barrier. Our aim is to incorporate dissipation in the framework of time dependent weak value and calculate the dwell time in that context. Here dissipation means loss of energy of the tunneling entity to the atomic modes of the medium in the barrier region. So in this particular case some of the energy of the tunneling entity is absorbed by the interacting medium of the barrier region. We consider time dependent Quantum weak value of certain operator as described by Davies \cite{7}. Then we will include dissipative interaction via the decay constant using the method of retarded Schr\"{o}dinger equation developed by Caldirola and Montaldi \cite{8} and arrive at an expression of finite non-zero dwell time.
To start with, the concept of Dwell time is discussed within the context of weak measurement theory in section II. In section III we discuss the time dependent weak values for two level state system. Based on this framework we calculate the dwell time for dissipative environment in section IV and finally some concluding remarks are made in section V.

\section{Dwell time and weak measurement}
\noindent
\vskip5pt
One of the commonly cited problem of measuring how much time it takes a quantum particle to cross the potential barrier is the non-existence of quantum mechanical time operator. However, it is possible to construct an operator
\beq\label{2.0}
\Theta_{(0,L)} = \Theta(x)-\Theta(x-L)
\eeq
  where $\Theta(x)$ and $\Theta(x-L)$ represents Heaviside functions. This operator measures whether the particle is in the barrier region or not.  Such a projection operator is hermitian and corresponds to a physical observable. It has eigenvalues 1 for the region $0\leq x\leq L$ and 0 otherwise. It's expectation value simply measures the integrated probability density over the region of interest. It is the expectation value divided by the incident flux which is referred to as the dwell time \cite{1a,8.1}. Ideally, transmission and reflection times $\tau_T$ and $\tau_R$ would when weighted by the transmission and reflection probabilities $|T|^2 \mbox{and} |R|^2$, yield the dwell time
\begin{equation}\label{2.1}
\tau_D = |T|^2 \tau_T + |R|^2 \tau_R
\end{equation}
In the last two decades a new approach to measurement theory in quantum mechanics has been developed by Aharanov and co-workers \cite{3,4}. This approach of ``weak measurement" differs from the standard ``von Neumann measurement" \cite{8a} in that the interaction between the measuring apparatus and the measured system is too weak to trigger a collapse of the wave function. Although the individual weak measurement of an observable has no meaning, one can of course obtain the expectation value to any desired accuracy by averaging a sufficiently large number of such individual results. In the standard approach of quantum mechanics, measurement comprises a collapse of the wave function which occurs instantaneously \cite{8b}. Avoiding wave function collapse allows the simultaneous measurement of non-commuting observables, though no violation of the uncertainty principle occurs because the individual measurements of each observable are very imprecise. Moreover, since it allows the system to evolve after the measurement as if unperturbed, it is possible to define average of a quantity conditioned to a given final state of the system. So if we are interested in the duration of some process, we can correspond this to a typical weak measurement extended in time, i.e. the interaction between the measuring probe and the system is not impulsive, but has a finite duration. Steinberg has shown \cite{8c} that these features make weak measurement theory a very promising background for the study of tunneling time in quantum mechanics.

Let us now first turn our attention to the theory of weak measurement. Weak value theory is a special consequence of the time symmetric reformulation of quantum mechanics. Whereas standard quantum mechanics describes a quantum system at time t using a state evolving forward in time from the past to t, weak value theory also uses a second state evolving backward in time from the future to t using the notion of pre and post selection \cite{3,4,5}. Consider the system prepared in an initial state $|i\rangle$ at a given initial time. At a given final time, the system is found to be in a final state $|f\rangle$. This means that a measurement performed at a particular initial time, selects only the systems in the preselected state $|i\rangle$, performs the weak measurement and at the final time again measurement performed to test whether the system is in the postselected state $|f\rangle$. Measurement is nothing but interaction with a measuring device having a particular initial state (Usually the position representation of the device wave function is taken as a Gaussian). System is made to interact with a pointer degree of freedom $Q$, via the interaction Hamiltonian
\begin{equation}\label{2.2}
 H_{int} = g(t)PA
\end{equation}
where $P$ is the conjugate momentum variable to the pointer position $Q$. It is convenient to take the function $g(t)$ impulsive and $g_0=\int g(t)dt=1$; it is non-zero only in a small interval. The initial state of the pointer variable is described by the Gaussian wavefunction
\begin{equation}\label{2.3}
\Phi_i(Q)=(\Delta^2 \pi)^{-\frac{1}{4}} e^{-\frac{Q^2}{2\Delta^2}}
\end{equation}
and the initial state of the system is given by
\begin{equation}\label{2.4}
|\psi_i\rangle=\sum_k a_k|a_k\rangle
\end{equation}
The initial state of the system is the eigenvector of the observable $A$. Now after the measurement interaction, the composite state of the system and the measuring device is given by
\begin{equation}\label{2.5}
(\Delta^2 \pi)^{-\frac{1}{4}}\sum_k a_k|a_k\rangle e^{-\frac{(Q-a_k)^2}{2\Delta^2}}
\end{equation}
In an ideal measurement, the relative shifts corresponding to different eigenvalues of the observable $A$ are large compared with the initial uncertainty in the pointer's position (given by the width $\Delta$) and the resulting lack of overlap between the final states leads to the irreversible collapse between different eigenstates of $A$. It is then found very close to the position corresponding to a particular eigenvalue of $A$. In weak measurement, the initial position of the pointer has a large uncertainty (ie large $\Delta$), so that the overlap between the pointer states remain close to unity, and hence the measurement does not constitute a collapse. The uncertainty in position measurement is large means that the momentum is more or less well defined. So it does not impart an uncertain kick to the particle. The measurement is weak in the sense that it disturbs the state of the particle as little as possible between the state preparation and post selection. Since the spread is large, the inaccuracy in measurement has to be compensated by large statistics (by averaging over a sub-ensemble). For post selection of state $|f\rangle$ of the system, the pointer wave function at final time is given by
\begin{equation}\label{2.6}
\Phi_f(Q)= (\Delta^2 \pi)^{-\frac{1}{4}}\sum_k a_k\langle f|a_k\rangle e^{-\frac{(Q-a_k)^2}{2\Delta^2}}
\end{equation}
After some mathematical analysis \cite{2} we find that
\begin{equation}\label{2.7}
\Phi_f(Q)\approx (\Delta^2 \pi)^{-\frac{1}{4}} e^{-\frac{(Q-A_w)^2}{2\Delta^2}}
\end{equation}
where
\begin{equation}\label{2.8}
A_w=\frac{\langle \psi_f|A|\psi_i\rangle}{\langle \psi_f|\psi_i\rangle}
\end{equation}
is denoted as the weak value of the observable $A$. We must note that the expression of this weak value may in generally be complex. However, the physical significance of the real and imaginary part is quite clear. The real part of the weak value corresponds to the mean shift of the pointer position and the imaginary part constitutes a shift in the pointer momentum. So, even though the imaginary part carries important physical significance it does not play any part in the measurement outcome since it does not corresponds to spatial translation of the pointer. It is also worthwhile to mention that besides being in generally complex, the magnitude of the weak value can lie outside the range of the eigenvalues \cite{3}.

Now let us discuss the calculation of  dwell time on the basis of weak measurement. The time taken by a particle to traverse certain potential barrier is measured by a clock consisting of an auxiliary system which interacts weakly with the particle as long as it stays in a given region. Aharanov et.al \cite{2} considered the interaction Hamiltonian as
\beq\label{2.9}
H_{int}=P_m \Theta_{(0,L)}
\eeq
where $m$ is the degree of freedom.
\beq\label{2.10}
\Theta_{(0,L)}= \left \{  \begin{array}{ll}
                 1 & \mbox{if}~~~ 0<x<L\\
                 0 & \mbox{otherwise}
                 \end{array} \right.
\eeq
This is the projection operator as we discussed earlier. It is the effective form of the potential, seen by a particle in the $S_z$ state, in the Stern-Gerlach experiment where (0,L) is the region of magnetic field. We  obtain the dwell time by calculating the weak value of the projection operator $\Theta_{(0,L)}$.
Now the weak value of any operator $A$ is expressed by equation (\ref{2.8}). If we divide the measurement into many short ones ($\Delta t$)
$ A\simeq \sum_{j=-\infty}^{\infty} A_j  $, we get \cite{2}
$$ <A_j>^w=C\Delta t \frac{\langle\psi_f(j\Delta t)|A|\psi_i(j\Delta t)\rangle}{\langle\psi_f(j\Delta t)|\psi_i(j\Delta t)\rangle}  $$
C is an arbitrary constant and can be set as $\frac{1}{\Delta t}$. In the limit $\Delta t\rightarrow 0$,
\beq\label{2.11}
A_w=\frac{\int_{-\infty}^{\infty}\langle\psi_f(t)|A|\psi_i(t)\rangle dt}{\langle\psi_f(0)|\psi_i(0)\rangle}
\eeq
For $A=\Theta_{0,L}$, this formula leads
\beq\label{2.12}
<\tau>^w= \frac{\int_{-\infty}^{\infty}dt\int_0^L \psi_f^{*}(x,t) \psi_i(x,t)dx}{\int_{-\infty}^{\infty} \psi_f^{*}(x,0) \psi_i(x,0)dx}
\eeq

Aharanov et.al \cite{2}  argued that direct calculation of the dwell time can be made using equation (\ref{2.12}). It shows that it tends to zero in the low energy limit. So irrespective of the length of the barrier, the particle traverses it in no time.

\section{Time dependent weak values of a two state system}

Now let us concentrate on time dependent pre and post selected states with the emphasis on decay of excited states. Let us consider the time evolution of a quantum mechanical state as
\beq\label{3.1}
|\psi(t)\rangle= U(t-t_0)|\psi(t_0)\rangle
\eeq
where $U(t-t_0)=e^{-iH(t-t_0)}$ is the time evolution operator.\\
\noindent In the light of the time evolution, the weak value of an operator $A$ at a time t, $t_i<t<t_f$, preselected at $t_i$ and postselected at $t_f$, can be defined as \cite{7}
\beq\label{3.2}
A_w= \frac{\langle\psi_f|U^{\dagger}(t-t_f)AU(t-t_i)|\psi_i\rangle}{\langle\psi_f|U^{\dagger}(t-t_f)U(t-t_i)|\psi_i\rangle}
\eeq
\\
\noindent Let us consider an electron of charge $e$ at rest in a magnetic field $\textbf{B}$. The interaction Hamiltonian is
\beq\label{3.3}
H=-\mathbf{\mu}.\textbf{B}
\eeq
where
\beq\label{3.4}
\mathbf{\mu}=-\frac{e\hbar \textbf{S}}{m}
\eeq
and
\beq\label{3.5}
\textbf{S}=\frac{1}{2}(\sigma_x,\sigma_y,\sigma_z)
\eeq
$\sigma_i$ are the Pauli spin matrices. For simplicity, let us suppose the magnetic field lies in the z direction. Then the hamiltonian looks like
\beq\label{3.6}
H=\hbar \omega \sigma_z
\eeq
The time evolution operator in this case looks like
\beq\label{3.7}
U(t)= \left (  \begin{array}{ll}
                 e^{i\omega t/2} & 0\\
                 0 & e^{-i\omega t/2}
                 \end{array} \right)
\eeq
From this we can get
\beq\label{3.8}
UU^{\dagger}=U^{\dagger}U= I
\eeq
and
\beq\label{3.9}
U(t_1-t_2)U(t_2-t_3)=U(t_1-t_3)
\eeq
So the unitary and evolution property hold.\\
\noindent Consider that at an initial time $t_i$ the state is polarized in the positive $x$ direction. Then
\beq\label{3.9a}
|\psi_i\rangle= \frac{1}{\sqrt{2}}\left (\begin{array}{ll}
                         1\\
                         1
                 \end{array} \right)
\eeq
The projection operator onto the eigenstate (\ref{3.9a}) is
\beq\label{3.9b}
P_{+}= \frac{1}{\sqrt{2}}\left (  \begin{array}{ll}
                 1 & 1\\
                 1 & 1
                 \end{array} \right)
\eeq
\noindent Now come to the case of the decay of an excited state by considering an initial excited two level atom coupled to a bath of $2N$ number of other two level atoms initially in their ground states.Due to the interaction with the bath atoms the concerning system is loosing enery to the bath modes. Choosing the ground state energies of all atoms to coincide and set to be zero, and setting the excited states $E_n$ to satisfy the relation
\beq\label{3.10}
E_n-E_0=n\Delta E,~~~~~-N\leq n\leq N
\eeq
the excited states are shown to be equispaced and distributed symmetrically about the excited state of the reference atom, labeled by $n=0$. For the simplicity of the problem, it is assumed that the reference atom is equally coupled to each of the atoms of the bath and so the interaction is described by the real constant Hamiltonian $H$.\\
The Schr\"{o}dinger equation is equivalent to the coupled differential equations
\beq\label{3.11}
\dot{a_0}=-i\sum_n Ha_n e^{-in\Delta Et}
\eeq
\beq\label{3.12}
\dot{a_n}=-iHa_0 e^{in\Delta Et}
\eeq
where $a_n$ is the amplitude of the excited state and we set $\hbar=1$. According to Davies \cite{7} equation (\ref{3.11}) and (\ref{3.12}) can be solved exactly by the method of Laplace transformation. Without going into the details of the calculations, which can be found in \cite{7}, we find that
\beq\label{3.13}
a_0(t)= e^{-\gamma (t-t_i)}
\eeq
where $\gamma$ is the decay constant. We discuss about this decay constant much more elaborately in the next section. The evolution operators $U(t)$ can also be found. If we consider that one atom at a time of the bath is excited, the evolution operator of the relevant sub-space of the full Hilbert space of states will be a $(2N+1)\times(2N+1)$ dimensional matrix, the components of which may be calculated from the equations (\ref{3.11}) and (\ref{3.12}). From equation (\ref{3.13}) it can be found
\beq\label{3.14}
U_{00}=e^{-\gamma t}
\eeq
in the limit $\Delta E\rightarrow 0$. Using this limiting solution, from the equations (\ref{3.11}) and (\ref{3.12}) it is found that
\beq\label{3.15}
U_{n0}=iH\left[\frac{e^{-\gamma t+in\Delta Et}-1}{\gamma -in\Delta E}\right ]
\eeq
which is also in the limit $\Delta E\rightarrow 0$. Using the relation $U^{\dagger}(t)=U(-t)$, we get the time dependent weak value of an operator $A$ as
\beq\label{3.16}
A_w=\frac{\langle \psi_f| U(t_f-t)AU(t-t_i|\psi_i\rangle}{\langle \psi_f|U(t_f-t_i)|\psi_i\rangle}
\eeq
Consider the operator $A$ to be chosen as the projection operator $P_{+}$ onto the excited state at time t, given that it is pre-selected in the excited state at time $t_i$ and post-selected to have decayed at time $t_f$. Let the possible choice of the final state be
\beq\label{3.17}
|\psi_f\rangle=|\psi_k\rangle
\eeq
This corresponds to the scenario that the atom is in the ground state and a photon of energy $E_k=k\Delta E$ having been emitted. This emitted photon may be absorbed by the bath modes due to the presence of the coupling. It can be shown that after some simple calculations the weak value gives
\beq\label{3.18}
P_w=\frac{U_{k0}(t_f-t)U_{00}(t-t_i)}{U_{k0}(t_f-t_i)}
\eeq
Using (\ref{3.14}) and (\ref{3.15}) it can be shown
\beq\label{3.19}
P_w=e^{-\gamma(t-t_i)} \left[\frac{1-e^{-\gamma(t_f-t)+ik\Delta E(t_f-t)}}{1-e^{-\gamma(t_f-t_i)+ik\Delta E(t_f-t_i)}}\right]
\eeq
In case of $E_k=E_0$, equation (\ref{3.19}) reduces to the simple expression
\beq\label{3.20}
P_w=e^{-\gamma(t-t_i)} \left[\frac{1-e^{-\gamma(t_f-t)}}{1-e^{-\gamma(t_f-t_i)}}\right]
\eeq
This is for the state to which the system approaches asymptotically as $t\rightarrow \infty$. For the post selection of state at a finite time $t_f$, according to \cite{7}, equation (\ref{3.20}) changes as
\beq\label{3.21}
P_w= e^{-\gamma(t-t_i)} \left[\frac{1-e^{-2\gamma(t_f-t)}}{1-e^{-2\gamma(t_f-t_i)}}\right]
\eeq
Now if we divide the measurement into many short ones, as we have discussed in the previous section, in comparison with the equation (\ref{2.11}) the weak value gives
\beq\label{3.22}
P_w=\int_{t_i}^{t_f} e^{-\gamma(t-t_i)} \left[\frac{1-e^{-2\gamma(t_f-t)}}{1-e^{-2\gamma(t_f-t_i)}}\right]dt
\eeq
Since the pre-selection and post selection are done at $t_i$ and $t_f$ respectively, correspondingly the limits of the integration are taken so. Now consequently this projection operator $P_{+}$ can be understood as the projection operator $\Theta_{0,L}$ as described in the previous section. Then equation (\ref{3.22}) gives the weak value of the operator $\Theta_{0,L}$, which in turn gives us the weak value of dwell time of the particle in the region of the magnetic field. Understanding the integral of the weak survival probability as dwell time also conforms with understanding of Winful \cite{5c}. As we have mentioned in the introduction, group delay ($\tau^G$) is understood as the lifetime of the energy storage in the barrier region and it is directly related to the dwell time ($\tau^D$) with an additive self-interference term ($\tau^I$).
\beq\label{3.22a}
\tau^G=\tau^D+\tau^I
\eeq
When the reflectivity is high the incident pulse spends much of its time ''dwelling'' in front of the barrier as it interferes with itself during the tunneling process. This excess dwelling is interpreted as the self-interference delay. Winful successfully disentangled this term from the dwell time \cite{5f}. Now if the surrounding of the barrier is dispersionless, then the self-interference term vanishes, resulting in the equality of the group delay and dwell time \cite{5e1}. In that case the dwell time will give the lifetime of energy storage in the barrier region. In our case the barrier region is dissipative (absorptive). So the integrated weak survival probability will give us lifetime of the remaining unabsorbed energy leaking through the barrier. What is more in this version of dwell time is that it includes the history of the interaction with the environment through the coupling term $\gamma$ as stated previously as decay constant. Therefore
\beq\label{3.23}
\tau_w=P_w=\int_{t_i}^{t_f} e^{-\gamma(t-t_i)} \left[\frac{1-e^{-2\gamma(t_f-t)}}{1-e^{-2\gamma(t_f-t_i)}}\right]dt
\eeq
Now before calculating the dwell time explicitly, we want to investigate the decay constant $\gamma$ much more elaborately. Since this $\gamma$ represents the coupling between the bath modes and the concerning system, this is the signature of dissipation.

\section{Derivation of Dwell time in dissipative environment}
The approach we are discussing here to incorporate dissipation in the dynamics of quantum system, was developed by Caldirola and Montaldi \cite{8} and Caldirola \cite{10} introducing a discrete time parameter ($\delta$) that could, in principle, be calculated from the properties of environment such as it's temperature and composition. It is used to construct a retarded Schr\"{o}dinger equation describing the dynamics of the states in presence of environmentally induced dissipation, which is given by
\beq\label{4.1}
H|\psi\rangle=i\frac{[|\psi(t)\rangle-|\psi(t-\delta)\rangle]}{\delta}
\eeq
Expanding $|\psi(t-\delta)\rangle$ in Taylor series, equation(\ref{4.1}) can be written as
\beq\label{4.2}
H|\psi\rangle=i\frac{[1-e^{-\delta \frac{\partial}{\partial t}}]|\psi(t)\rangle}{\delta}
\eeq
Setting the trial solution as
\beq\label{4.3}
|\psi(t)\rangle=e^{-\alpha t}|\psi(0)\rangle
\eeq
we solve for $\alpha$ to get
\beq\label{4.4}
\alpha=\frac{1}{\tau}\ln(1+iH\delta)
\eeq
Substituting $\alpha$ in equation(\ref{4.2}) we find that even the ground state decays. To stabilize the ground state, Caldirola and Montaldi \cite{8} rewrite equation (\ref{4.1}) as
\beq\label{4.5}
(H-H_0)|\psi\rangle=i\frac{[|\psi(t)\rangle-|\psi(t-\delta)\rangle]}{\delta}
\eeq
Where $H_0$ represents the ground state. In this case we get
\beq\label{4.6}
\alpha=\frac{1}{\tau}\ln\left(1+i(H-H_0)\delta\right)
\eeq
For a spin half system in a magnetic field ($\mathbf{B}$), the Hamiltonian is
\beq\label{4.7}
H=\frac{e}{m}S_z B
\eeq
For the eigenvalues of (\ref{4.7}) we have
\beq\label{4.8}
E_{+}=\frac{eB}{2m},~~|\psi\rangle=\left (\begin{array}{ll}
                         1\\
                         0
                 \end{array} \right)
\eeq
\beq\label{4.9}
E_{-}=-\frac{eB}{2m},~~|\psi\rangle=\left (\begin{array}{ll}
                         0\\
                         1
                 \end{array} \right)
\eeq
Now let us take the state to be initially polarized in the x direction as shown by equation (\ref{3.9a}). Now following Wolf \cite{11}, to generate the states at time t, we use
\beq\label{4.10}
|\psi(t)\rangle=\exp\left[-\frac{t}{\delta}\ln(1+i(H-H_0)\delta)\right]|\psi(0)\rangle
\eeq
where
\beq\label{4.11}
H_0=-\frac{eB}{2m}  \left (  \begin{array}{ll}
                 1 & 0\\
                 0 & 1
                 \end{array} \right)
\eeq
and
\beq\label{4.12}
H=\frac{eB}{2m}\left (  \begin{array}{ll}
                 1 & ~~ 0\\
                 0 & -1
                 \end{array} \right)
\eeq
Therefore
\beq\label{4.13}
H-H_0=\frac{eB}{m}\left (  \begin{array}{ll}
                 1 & 0\\
                 0 & 0
                 \end{array} \right)
\eeq
So we find the state at time t as
\beq\label{4.14}
|\psi(t)\rangle= \frac{1}{\sqrt{2}}\left (\begin{array}{ll}
                         \exp\left[-\frac{t}{\delta}\ln(1+\frac{ieB\delta}{m})\right]\\
                         ~~~~~~~~1
                 \end{array} \right)
\eeq
Expanding the logarithmic term upto 3rd order
\beq\label{4.15}
\ln\left(1+\frac{ieB\delta}{m}\right)=\frac{ieB\delta}{m}+\frac{e^2B^2\delta^2}{2m^2}-\frac{ie^3B^3\delta^3}{3m^3}
\eeq
the time evolution takes the form
\beq\label{4.16}
\exp\left[-i\left(\frac{eB}{m}-\frac{e^3B^3\delta^2}{3m^3}\right)t-\frac{e^2B^2\delta}{2m^2}t\right]
\eeq
From (\ref{4.16}) we find the modified precession frequency
\beq\label{4.17}
\omega'=2\omega \left(1 - \frac{e^2B^2\delta^2}{3m^2}\right)
\eeq
where $\omega=\frac{eB}{2m}$ is the unmodified precession frequency.
We also find the decay rate as
\beq\label{4.18}
\gamma=\frac{e^2B^2\delta}{2m^2}
\eeq
Again from equation (\ref{4.17}) we find the time scale $\delta$ as
\beq\label{4.18a}
\delta=\frac{1}{2\omega}\sqrt{3\left(1-\frac{\omega'}{2\omega}\right)}
\eeq
So the decay constant takes the form
\beq\label{4.18b}
\gamma=\omega\sqrt{3\left(1-\frac{\omega'}{2\omega}\right)}
\eeq

Putting this in equation (\ref{3.23}) and integrating, we find the dwell time as
\beq\label{4.19}
\tau_w=\frac{1}{\omega\sqrt{3\left(1-\frac{\omega'}{2\omega}\right)}}\coth\left(\frac{\omega T}{2}\sqrt{3\left(1-\frac{\omega'}{2\omega}\right)}\right)
\eeq
where $T=t_f-t_i$.
Here we arrive at the expression of dwell time for a spin half particle traversing through a magnetic potential barrier in presence of dissipation, which was pre-selected in a state with energy $\omega$ at an initial time $t_i$ and post-selected in a state with energy $\omega'$ at a final time $t_f$. Let us consider a particular case where the particle is pre-selected in the spin up state with $\omega=\omega_{+}=\frac{eB}{2m}$ and post-selected in the spin down state with $\omega'=\omega_{-}=-\frac{eB}{2m}$. Then the dwell time takes the form
\beq\label{4.20}
\tau_w=\frac{\sqrt{2}}{3\omega}\coth\left(\frac{3\omega T}{2\sqrt{2}}\right)
\eeq
Similarly we can find the dwell time for a spin half particle traversing through similar kind of barrier, pre-selected and also post-selected in the spin up state. If we consider the interpretation of dwell time as stated by Winful, these finite non-zero dwell times will give us the life time of the decaying states in the magnetic barrier region, depending on the pre and post selection of the states.
\vspace{.4cm}
\section{Conclusion}
\vspace{-.3cm}
From the result of our calculation, it is evident that presence of dissipative interaction with the bath modes precludes the zero time tunneling, i.e. the instantaneous release of energy. This can be explained in terms of an efficient energy transfer from the particle motion to the environmental modes, and loss of ''memory'' of the original tunneling direction. Here the barrier is acting as a lumped capacitor with coupling to the environment providing
an effective dissipation.The delay is caused by the energy storage in the barrier region. It is also worth to mention here that a connection between dwell time and realistic examples of lifetime of decaying states was established in some works of Kelkar et.at \cite{12,13}, where they have dealt with the phenomena of Alpha decay by investigating the quantum time scales of tunneling. They have formulated the half-life of the decaying state in terms of the transmission dwell time. They found that the major bulk of the half-life of a medium or super heavy radioactive
nucleus is spent in front of the barrier before tunneling.The time spend inside the barrier is much smaller, though it is not negligible. Now if the barrier is considered to be dissipative, we can heuristically argue that due to the energy transfer to the nuclear modes will result in the loss of memory of original tunneling direction and hence cause an extra delay, enhancing the dwell time inside the barrier. The alpha particles or clusters formed inside the nucleus can undergo some restructuring within the nucleus. Work has been done on the derivation of necessary formulae accounting for the final state interaction between the knocked out cluster and the residual nucleus \cite{14}. This kind of interaction may account for the dissipation. Admitting this to be a heuristic argument on the subject of alpha decay, we can consider it to be an important subject for a future work.


\begin{thebibliography}{}
\bibitem{1}M.D. Stenner,D.J.Gauthier and M.A.Neifeld(2003); Nature(London) {\bf 425},695.
\bibitem{1a} E.H.Hauge and J.A.St{\o}veng(1989); Rev.Mod.Phys. {\bf 61},917.
\bibitem{2}Y.Aharonov,N.Erez and B.Reznik(2003); Journal of Modern Optics {\bf 50},1139.
\bibitem{2a}D. Sokolovski,A. Z. Msezane and V. R. Shaginyan(2005); Phys.Rev.A {\bf 71},064103.
\bibitem{3}Y.Aharonov,D.Albert and L.Vaidman(1988); Phys.Rev.Lett. {\bf 60},1351.
\bibitem{4}Y.Aharonov and L.Vaidman(1990); Phys.Rev.A {\bf 41},11.
\bibitem{5}Y.Aharonov,D.Albert,A.Casher and L.Vaidman(1986); \textit{New Techniques and ideas in quantum measurement theory}(New York: N.Y. Academy of Science), p 417.
\bibitem{5a}A. M. Steinberg, P. G. Kwiat, and R. Y. Chiao(1993); Phys. Rev. Lett. {\bf 71}, 708.
\bibitem{5b}Ch. Spielmann, R. Szip\"{o}cs, A. Stingl, and F. Krausz(1994), Phys. Rev. Lett. {\bf 73}, 2308.
\bibitem{5c}H.G.Winful(2006); New Journal of Physics {\bf 8},101.
\bibitem{5d}H.G.Winful(2002); Opt.Express {\bf 10},1491.
\bibitem{5e}H.G.Winful(2003); IEEE J. Sel. Top. Quantum Electron {\bf 9},17.
\bibitem{5e1}H.G.Winful(2003);Phys. Rev. E {\bf 68}, 016615.
\bibitem{5e2}H.G.Winful(2006);Phys. Rep. {\bf 436}, 1.
\bibitem{5f}H.G.Winful(2003); Phys. Rev. Lett. {\bf 91}, 260401.
\bibitem{7}P.C.W.Davies(2009); Phys. Rev. A {\bf 79}, 032103.
\bibitem{8}P.Caldirola and M.Montaldi(1979); Nouvo Cimento Soc. Ital. Fis.B {\bf 53},291.
\bibitem{8.1}M.B\"{u}ttiker(1983); Phys.Rev.B {\bf 27},6178.
\bibitem{8a}J. von Neumann(1955); \textit{Mathematical foundation of quantum mechanics}, Princeton University Press, Princeton.
\bibitem{8b}W. Heisenberg(1930); \textit{The physical principles of quantum theory}, Dover, Newyork, p. 39.
\bibitem{8c}A.M. Steinberg(1995); Phys.Rev. A {\bf 52}, 32.
\bibitem{10}P.Caldirola(1976); Lett. Nouvo Cimento Soc. Ital. Fis. {\bf 16},151.
\bibitem{11}C.Wolf(1999); Can.J.Phys. {\bf 77},785.
\bibitem{12}N.G.Kelkar,H.M.Castaneda and M.Nowakowski(2009); Europhysics Lett. {\bf 85},20006.
\bibitem{13}N.G.Kelkar(2007); Phys.Rev.Lett. {\bf 99},210403.
\bibitem{14}A. Sakharuk, V. Zelevinsky and V. G. Neudatchin(1999); Phys. Rev. C {\bf 60}, 014605.
\end{thebibliography}
\end{document}